# Simulation of Optimum values of device parameters to attain Peak to Valley current Ratio (PVCR) in Resonant tunneling diodes (RTD)

*Sushree Ipsita[1], Prasanta Kumar Mahapatra[1], Pradipta Panchadhyayee[2]

[1]*Department of Physics, Siksha 'O' Anusandhan (Deemed to be) University, Khandagiri, Bhubaneswar-751030, Odisha, India*

[2]*P.K. College, Contai, Purba Medinipur, West Bengal-721404, India*

**Abstract**: Relations for the optimum well width, barrier width and width of the spacer layer to achieve highest PVCR on the basis of effective mass and barrier height in RTDs is proposed. The optimum spacer layer is found to be half of the de-Broglie wavelength associated with the bound state of the corresponding finite quantum well. The proposed relations for the optimum parameters can be used to design RTD based on any two appropriate materials to attain highest PVCR. The effect of doping concentrations on PVCR and peak current was studied. As case study, we have considered the GaAs/Ga$_{0.7}$Al$_{0.3}$As and GaN/Ga$_{0.7}$Al$_{0.3}$N RTDs. The current density obtained using the tunneling coefficient based on transfer matrix approach takes in to account the variation in the electric field in the well and barrier region on account of variation in the dielectric constant in the material.

## Introduction

The chip density and operating speed of Si based ICs, with continuous scaling of device dimensions, have registered a growth in consonance with Moore's law [1]. With the device density approaching to ten billion devices, the device size becomes comparable to the electron De Broglie wave length in the order of 50 nm. The performance of the devices in the nanometer order is strongly affected by the quantum effects and thus necessitated a search for alternative devices which encompass the quantum effect. Resonant tunneling diode based on quantum tunneling through two or more very thin barrier layers is one such alternative device [2], which has received considerable attention owing to the appearance of negative differential conductance (NDC) over narrow voltage ranges in its I-V characteristics [3]. The Resonant Tunneling Diodes (RTD) based on double barrier semiconductor systems (DBS), having THz operational capabilities and low voltage operation [4, 5], and have potential applications in high frequency oscillators [6, 7, and 8], fast digital switches [9, 10], photo-detectors [11] and quantum integrated circuits [12 and 13].

In multi-barrier systems (MBS) for certain incident energies below the barrier height, the transmission coefficient becomes unity and the MBS becomes totally transparent for electrons with these energies [14]. The complete transmission across the system is termed as "Resonant Tunneling" and the energies corresponding to the resonant tunneling is coined as quasi resonant tunneling energy states. With the application of the field, the current–voltage characteristics in MBS exhibit NDC regimes with resonant type anomalies similar to those predicted in single crystals in the Wannier Stark Ladder regime [15, 16]. Esaki and Tsu in 1974 first experimentally demonstrated the resonant peaks and negative differential conducting regions in the tunneling current in the $Al_xGa_{1-x}As/GaAs/Al_xGa_{1-x}As$ DBS and explained the observed features on the basis of tunneling from the contact layer to the quantized levels in the GaAs well layer[17,18]. It is being explained that the current starts to increase as this stark shifted level approaches the Fermi level and attains a peak value when the energy level coincides with the conduction band minima of the contact layer. With further increase in the field, there is a sharp decrease in the current and reaches a minimum known as the valley current and then the process gets repeated as the second quasi resonant energy level approaches the Fermi level. The ratio of the peak current to that of valley current in the device popularly known as peak to valley current ratio (PVCR), is a vital parameter for device performance [19] in the digital circuits.

The technological importance of the RTDs was recognized in 1983 with the findings of Sollner that the intrinsic charge transport in a RTD can respond to change in voltages in a time range of 0.1ps or less, opening its application potentiality in electronic devices operating at Terahertz frequencies [20]. RTDs have been reported to achieve a frequency of 2.2THz as opposed to 215GHz realized in conventional complementary metal oxide semiconductor (CMOS) transistors [21]. The very high switching speeds of the RTDs have found potential applications in high resolution radar and imaging systems in environments of low visibility [22, 23].

The Si/SiGe RTDs, which are easy to fabricate using CMOS technology, have low PVCR with reported values of 2.9 and a peak current density (PCD) of 4.3 kA/cm$^2$ owing to the short barrier height [24]. The maximum PVCR of 21.7 and 35 have been reported for GaAs/AlAs [25] and InGaAs/AlAs [26] RTDs respectively. For high frequency operation an improvement in power output expressed as $\Delta I.\Delta V$ need an increase in the peak current especially at low voltage operating condition.

Keeping in mind that the PVCR is the most important factor for application in digital circuit [27, and 28], here we aim to find the optimum fabrication conditions to attain highest PVCR in the RTD. To the best of our knowledge we have not come across any systematic study on the optimization of device parameters for achieving highest PVCR. For the purpose of finding the optimum conditions for RTDs, we consider two different systems GaAs/AlGaAs and GaN/AlGaN. The material properties in the RTD appear through the conduction band offset at the hetero-junction and the effective mass of the electrons in the well and barrier materials. A general relation for the optimum device parameters (well width, barrier width and spacer layer) for achieving the highest PVCR based on the effective mass of electrons and the barrier height characteristic to the well and barrier material is presented. The relation will provide the experimentalist a tool to calculate the optimum device parameters for RTDs based on any two materials.

**Theoretical development**

A schematic layer structure of the RTD adopted for the study of the current ~ voltage relation is presented in Fig. 1. The layout consists of a low-gap material A of thickness '$a$' (the well) sandwiched between two layers of the high-gap material B of thickness '$b$' (barrier). Two spacer layers of the low gap material A, with thickness '$w$' are grown on either side of the DBS followed by heavily doped contact layers of the material A. The DBS potential profile appears as alternate rectangular wells and barriers at the conduction band edge along the growth direction and is considered to be superimposed on the intrinsic periodic potential of the host materials A and B. The inherent lattice potentials of the well and barrier layers are taken care through the corresponding effective mass $m_w^*$, and $m_b^*$ respectively, obtained on the basis of $\varepsilon$-$k$ relation of the materials. The height of the potential barrier has been taken as $V_0$, which is equal to the band mismatch at the conduction band edge, that can be estimated using various theoretical models and experimental measurements and is proposed to be 88% of the difference between the energy gap of the two materials [29]. The effect of the superposing potential is considered in the envelop function approach.

The current density for the RTD is computed on the basis of the integration of the transmission coefficient across the structure over the energy distribution of the electrons based on the Fermi energy using a relation proposed by Tsu [17]. The transmission coefficient across the electrically biased RTD structure is obtained through transfer matrix approach using the exact Airy function formalism and effective mass-dependent boundary condition [30]. In our approach, the variation in the electric field in the well and barrier region on account of variation in the dielectric constant is incorporated in the transmission coefficient calculation. All previous calculations have considered a uniform electric field throughout the specimen. For a comprehensive analysis of the aspects related to the I-V characteristics of the RTD, we need to know the energy levels associated with the quantum well, the transmission coefficient, resonant tunneling energy in the corresponding DBS both under unbiased and electrically biased conditions. In Appendix A, we have provided briefly the procedure to determine the quantum well energy and the corresponding wave function.

**Transmission Coefficient in biased RTD structure**

In Fig. 2(a) we have presented the potential profile in black solid line for the diode in which the electrical potential $V_a$ is applied between the points 0 and L. With application of the potential $V_a$ across the device, the potential profile of the biased MBS along the growth direction appears as:

$$V(z) = \begin{cases} 0 & for\ x < x_1 \\ -eE_w x & for\ x_{2n-1} \leq x < x_{2n};\ n = 1,2,3 \\ V_0 - eE_B x & for\ x_{2n} \leq x < x_{2n+1};\ n = 1,2 \\ -eV_a & for\ x \geq x_6 \end{cases} \quad (1)$$

where $E_w = \frac{V_a}{\kappa_w}\left(\frac{L_w}{\kappa_w} + \frac{L_B}{\kappa_B}\right)^{-1}$ and $E_B = \frac{V_a}{\kappa_B}\left(\frac{L_w}{\kappa_w} + \frac{L_B}{\kappa_B}\right)^{-1}$ represent the electric field in the well and barrier material respectively; $L_w = w_L + a + w_R$ and $L_B = 2b$ representing the total width of well and barrier materials respectively; $\kappa_w$ and $\kappa_B$ represent the dielectric constants in the two materials; the six points $x_i$ with $i$ = 1 to 6 represent the points of slope change or discontinuity in the potential profile and have the values $x_1 = 0$, $x_{2n} = w_L + (n-1)(a+b)$, $x_{2n+1} = x_{2n} + b$; n = 1,2 and $x_6 = L = L_w + L_B$. The envelope function corresponding to the different regions now appears as

$$\begin{cases} \Psi_0(x) = A_0 e^{ikx} + B_0 e^{-ikx} & for\ x < x_1 \\ \psi_{2n-1}(x) = A_{2n-1} Ai(\eta) + B_{2n-1} Bi(\eta); & x_{2n-1} \leq x < x_{2n};\ n = 1,2,3 \\ \psi_{2n}(x) = A_{2n} Ai(\xi) + B_{2n} Bi(\xi), & x_{2n} \leq x <;\ n = 1,2 \\ \Psi_6(x) = A_6 e^{ik'x} + B_6 e^{-ik'x} & for\ x > x_6 \end{cases} \quad (2)$$

where, $k = \frac{\sqrt{2m_w^* \varepsilon}}{\hbar}$ ; $\eta = -\alpha_1 x - \lambda_1$ , $\alpha_1^3 = \frac{2m_w^* e E_w}{\hbar^2}$, $\lambda_1 = \frac{2m_w^* \varepsilon}{\alpha_1^2 \hbar^2}$ ;

$\xi = -\alpha_2 x - \lambda_2$ , $\alpha_2^3 = \frac{2m_B^* e E_B}{\hbar^2}$, $\lambda_2 = \frac{2m_B^*(V_0 - \varepsilon)}{\alpha_2^2 \hbar^2}$; and $k' = \frac{\sqrt{2m_w^*(\varepsilon + eV_a)}}{\hbar}$;

*Ai* and *Bi* being Airy functions of the first and second kind.

Using effective mass based boundary condition at each of the junctions the coefficients of the outgoing wave $A_6$ and $B_6$ can be obtained from the initial coefficients $A_0$ and $B_0$ through the transfer matrix $T_E$ as:

$$\begin{bmatrix} A_6 \\ B_6 \end{bmatrix} = T_E \begin{bmatrix} A_0 \\ B_0 \end{bmatrix} \tag{3}$$

where, $T_E = \prod_1^6 M_{En}$ And $\begin{bmatrix} A_n \\ B_n \end{bmatrix} = M_{En} \begin{bmatrix} A_{n-1} \\ B_{n-1} \end{bmatrix}$

The six transfer matrices appear as:

$$M_{E1} = \frac{1}{w} \begin{bmatrix} Bi'(\eta_1) + \frac{ik}{\alpha_1} Bi(\eta_1) & Bi'(\eta_1) - \frac{ik}{\alpha_1} Bi(\eta_1) \\ -\left( Ai'(\eta_1) + \frac{ik}{\alpha_1} Ai(\eta_1) \right) & -\left( Ai'(\eta_1) - \frac{ik}{\alpha_1} Ai(\eta_1) \right) \end{bmatrix}$$

$$M_{E\,2n} = \frac{1}{w} \begin{bmatrix} \begin{vmatrix} Ai(\eta_{2n}) & Bi(\xi_{2n}) \\ \mu\, Ai'(\eta_{2n}) & Bi'(\xi_{2n}) \end{vmatrix} & \begin{vmatrix} Bi(\eta_{2n}) & Bi(\xi_{2n}) \\ \mu\, Bi'(\eta_{2n}) & Bi'(\xi_{2n}) \end{vmatrix} \\ -\begin{vmatrix} Ai(\eta_{2n}) & Ai(\xi_{2n}) \\ \mu Ai'(\eta_{2n}) & Ai'(\xi_{2n}) \end{vmatrix} & -\begin{vmatrix} Bi(\eta_{2n}) & Ai(\xi_{2n}) \\ \mu\, Bi'(\eta_{2n}) & Ai'(\xi_{2n}) \end{vmatrix} \end{bmatrix}; \quad n=1, 2$$

$$M_{E\,2n+1} = \frac{1}{w} \begin{bmatrix} \begin{vmatrix} Ai(\xi_{2n+1}) & Bi(\eta_{2n+1}) \\ \mu^{-1} Ai'(\xi_{2n+1}) & Bi'(\eta_{2n+1}) \end{vmatrix} & \begin{vmatrix} Bi(\xi_{2n+1}) & Bi(\eta_{2n+1}) \\ \mu^{-1} Bi'(\xi_{2n+1}) & Bi'(\eta_{2n+1}) \end{vmatrix} \\ -\begin{vmatrix} Ai(\xi_{2n+1}) & Ai(\eta_{2n+1}) \\ \mu^{-1} Ai'(\xi_{2n+1}) & Ai'(\eta_{2n+1}) \end{vmatrix} & -\begin{vmatrix} Bi(\xi_{2n+1}) & Ai(\eta_{2n+1}) \\ \mu^{-1} Bi'(\xi_{2n+1}) & Ai'(\eta_{2n+1}) \end{vmatrix} \end{bmatrix}; \quad n=1, 2$$

$$M_{E6} = \frac{1}{2} \begin{bmatrix} Ai(\eta_6) + \frac{i\alpha_1}{k'} Ai'(\eta_6) e^{-ik'L} & Bi(\eta_6) + \frac{i\alpha_1}{k'} Bi'(\eta_6) e^{-ik'L} \\ Ai(\eta_6) - \frac{i\alpha_1}{k'} Ai'(\eta_6) e^{ik'L} & Bi(\eta_6) - \frac{i\alpha_1}{k'} Bi'(\eta_6) e^{ik'L} \end{bmatrix} \tag{4}$$

where $\mu = \left( \frac{m_w^*}{m_B^*} \right)^{2/3}$, $\eta_n = -\alpha_1 x_n - \lambda_1$, $\xi_n = -\alpha_2 x_n - \lambda_2$ and $w$ is the Wronskian of the Airy functions. Using the condition that there is no reflected component beyond the point $x=L$, the transmission coefficient $T_{cE}$ for an electron with incident energy, $\varepsilon$, across the biased RTD can be obtained as [16] :

$$T_{cE}(\varepsilon) = \frac{|A_6|^2}{|A_0|^2} = \frac{|Det[T_E]|^2}{|(T_E)_{22}|^2} \tag{5}$$

The transmission coefficient for the DBS is computed on the basis of (5) for incident energy up to 1eV.

**Tunneling Current Density**

The current density J is determined [17] using the transmission coefficient from the relation

$$J = \frac{e m_w^* k_B T}{2\pi^2 \hbar^3} \int \frac{k'}{k} T_{cE}(\varepsilon) \ln \left[ \frac{1 + \mathrm{Exp}((\varepsilon_F - \varepsilon)/k_B T)}{1 + \mathrm{Exp}((\varepsilon_F - \varepsilon - eV_a)/k_B T)} \right] d\varepsilon \tag{6}$$

where $\varepsilon_F$, $\varepsilon$, $V_a$ and $T$ represent the Fermi energy, the incident energy of the electron, the applied bias and the temperature in absolute scale respectively.

The tunneling current density is computed on the basis of (6) for the RTD using the MATHEMATICA software for a wide spectrum of parameters like, well width, barrier width and height, effective mass of the materials, width of spacer layer and the doping concentration in the contact layers. The optimum device parameters are obtained for achieving the maximum PVCR for different systems. The effect of thermal energy and doping concentration on the voltage corresponding to Peak current is ascertained.

**Numerical Analyses**:

Our numerical analysis is mostly concerned with (a) Evaluating the current density vs. applied voltage in the RTD on the basis of (6) and determining peak and valley current, PVCR and width of the NDR region in I-V graph, (b)Finding the optimized well, barrier and spacer layer widths for the systems to achieve highest PVCR, (c) Finding a relation for the optimum device parameters as a function of barrier height and effective mass of well and barrier materials, (d) Study the variation of peak current with Fermi energy which depends on the doping concentration and the dopant ionization energy, (e) Finding the life time corresponding to the optimum state which have a say on the operating frequency. Here we have considered two different systems, GaAs/Ga$_{0.7}$Al$_{0.3}$As and GaN/Ga$_{0.7}$Al$_{0.3}$N, to study the above mentioned points. The necessary parameters used for the transmission coefficients are given in Table. I.

**Table. I:** Required parameters for GaAs/Ga$_{0.7}$Al$_{0.3}$As and GaN/Ga$_{0.7}$Al$_{0.3}$N

| System | $\varepsilon_{gw}$ (eV) | $\varepsilon_{gb}$ (eV) | $V_0$ (eV) | $m_w^*$ ($m_0$) | $m_b^*$ ($m_0$) |
|---|---|---|---|---|---|
| GaAs/Ga$_{0.7}$Al$_{0.3}$As | 1.4240 | 1.8476 | 0.3728 | 0.067 | 0.0904 |
| GaN/Ga$_{0.7}$Al$_{0.3}$N | 3.3900 | 3.9708 | 0.5111 | 0.200 | 0.2343 |

The relations used for the determination of the band gap, effective mass of the conduction electrons in the mixed compound and the conduction band offset at the hetero-junction of well and barrier is detailed in Appendix-B.

We have computed the current density at room temperature for the GaAs /Ga$_{0.7}$Al$_{0.3}$As and GaN /Ga$_{0.7}$Al$_{0.3}$N RTD systems for various values of doping concentration. It is worth pointing here that the doping concentrations enter the current density calculation of (6) through the parameter $\varepsilon_F$, the Fermi energy. The relation for computing the Fermi energy from doping concentrations and dopant ionization energy is outlined in Appendix-B.

**Results and discussion**

In the endeavor to find the device specifications for the maximum peak to valley ratio for different RTD systems, we first present the numerical results as regard to the current density ~ applied potential for the GaAs –Ga$_{0.7}$Al$_{0.3}$As RTD system computed on the basis of (6) in Figs. 3 and 4. The J-V characteristics is calculated for a temperature of 300K, taking the system parameter $\varepsilon_f = 0$ i.e. the Fermi level coinciding with the conduction band edge of GaAs. As mentioned in appendix –B, this value of $\varepsilon_f$ corresponds to a doping of 0.0053% in the contact layer.

**Effect of barrier width on the PVCR for GaAs –Ga$_{0.7}$Al$_{0.3}$As RTD**

In Fig. 3(a), we demonstrate the variation in the J ~ V curve with respect to the change in barrier width from 43 Å to 51 Å keeping the well width constant at 38 Å and the spacer layer at 19 Å. For better comprehension, the data pertaining to the Fig. 3(a) are tabulated in Table II. The table shows that with the increase in barrier width, the peak current, valley current and the width of the NDC region decrease. However, for a constant well width in the system, with an increase in barrier width, the PVCR first increases, attains a maximum value and then decreases. In fact, for the GaAs –Ga$_{0.7}$Al$_{0.3}$As system of well width of 38 Å, the PVCR gets maximized for a barrier width of 47 Å. The findings imply the existence of a particular barrier width corresponding to each well width for which the PVCR is maximized.

To find the reason behind the existence of the optimum barrier width corresponding to a finite well width, we computed the resonant tunneling energy in the RTD structure for barrier width in the range 20Å to 75Å keeping the well width at 38Å. The transmission coefficient and the resonant tunneling energy for various values of the barrier width in the GaAs –Ga$_{0.7}$Al$_{0.3}$As RTD structure keeping the well width at 38Å is presented in Fig.3(b) and 3(c) respectively.

Table. II: Variation of PVCR and other parameters with barrier width as depicted in Fig. 3

| Well width(Å) | Barrier width (Å) | $V_P$ (Volt) | Log ($J_P$) (A/cm$^2$) | $V_V$ (Volt) | Log($J_V$) (A/cm$^2$) | PVCR | Width of NDC (V) |
|---|---|---|---|---|---|---|---|
| 38 | 43 | 0.2155 | 3.6183 | 0.2754 | 1.87769 | 55.0 | 0.0599 |
| 38 | 47 | 0.2149 | 3.3288 | 0.26575 | 1.55799 | 59.0 | 0.05085 |
| 38 | 51 | 0.21425 | 3.0389 | 0.2583 | 1.29050 | 56.0 | 0.04405 |

As can be seen in these figures, the resonant tunneling energy increases with increase in barrier width up to 47Å and then remains constant at 0.12349eV with further increase in barrier width. It is worth pointing here that the bound state energy of the finite quantum well formed by $Ga_{0.7}Al_{0.3}As$-38ÅGaAs –$Ga_{0.7}Al_{0.3}As$ is equal to 0.12349eV (as shown in Appendix-A). Thus the optimum barrier width corresponding to a finite well width correspond to the condition of minimum width of the barrier at which resonant tunneling energy in the RTD structure equals the bound state energy of the corresponding finite quantum well. With further increase in barrier width the PVCR decreases as tunneling probability decreases.

**Effect of well width on the PVCR GaAs –$Ga_{0.7}Al_{0.3}As$ RTD**

The results in the last subsection motivated us to find the optimum barrier width which gives maximum PVR for different well widths and the results for GaAs –$Ga_{0.7}Al_{0.3}As$ RTD system are presented in Fig. 4(a). In all these calculations the spacer layer is maintained at half of the well width. As is evident from the figure, for maximum PVCR, with an increase in well width, the barrier width is first found to increase, attains a peak value of around 53 Å corresponding to the well width in the region of 25 Å to 28 Å and then decreases with an increase in well width.

To study the variation of PVCR with well width we present in Fig. 4(b) the variation of J ~ V for three well widths 34Å, 38 Å and 42 Å. For these well widths, we consider the barrier widths for which the PVCR is maximized and is equal to 50Å, 47Å and 45Å respectively. The data corresponding to the Fig. 4(b) are presented in Table III. As can be seen from the tabular data, with an increase in well width, the PVCR first increases, attains a maximum value and then decreases. The PVCR of the device having well width of 38 Å, and 47 Å is found to be higher than that corresponding to the well width of 34 Å and 42 Å. In Fig.4(c), the variation of maximum PVCR corresponding to various well widths in the range of 10Å to 50Å for the GaAs –$Ga_{0.7}Al_{0.3}As$ RTD is presented. The graph clearly shows that in the GaAs –

$Ga_{0.7}Al_{0.3}As$ based RTD the highest PVCR correspond to the device with well and barrier width of 38 Å and 47 Å respectively.

**Table III.** Variation of PVCR and other parameters with well width corresponding to the barrier width which provides maximum PVCR. Data taken from Fig. 4(b).

| Well width(Å) | Barrier width (Å) | $V_P$ (Volt) | Log ($J_P$) (A/cm²) | $V_V$ (Volt) | Log($J_V$) (A/cm²) | PVCR | Width of NDC |
|---|---|---|---|---|---|---|---|
| 34 | 50 | 0.2438 | 3.22114 | 0.2962 | 1.4774 | 55.4 | 0.0524 |
| 38 | 47 | 0.2149 | 3.32883 | 0.26575 | 1.55799 | 59.0 | 0.05085 |
| 42 | 45 | 0.19025 | 3.37451 | 0.2374 | 1.62682 | 55.9 | 0.04715 |

Further with an increase in well width, the peak current and valley current increase while the peak voltage decrease. The decrease in the peak voltage can be explained on the basis of decrease in the bound state energy level in the corresponding quantum well and the resonant tunneling energy in the DBS with wider wells.

**Effect of spacer layer on the PVCR GaAs – $Ga_{0.7}Al_{0.3}As$ RTD**

In Fig. 5, we present the variation of PVCR with the width of the spacer layer for the GaAs – $Ga_{0.7}Al_{0.3}As$ RTD with the well width of 38 Å and barrier width of 47 Å. The tunneling current density and the PVCR is found to increase with increase in spacer layer width up to 62Å and then decreases. To analyze the reasons behind such a variation, we would like to point that for the finite $Ga_{0.7}Al_{0.3}As$-38ÅGaAs –$Ga_{0.7}Al_{0.3}As$ well, the bound state energy amounts to 0.12349eV and the corresponding De Broglie wave length is estimated at 135Å. In the spacer layer, the electron wave consists of the incident wave with probability density $|A|^2$ (considered as unity) and a reflected wave whose probability density $|B|^2$(varies with the incident energy and is found to be very small at resonant condition). However, these incident and reflected waves with same de Broglie wave length overlap to produce some sort of a travelling wave super imposed by a standing wave of small amplitude $2|A||B|$. The standing wave have anti nodes separated by 67.5Å with an initial phase of $\tan^{-1}\left(\frac{Im(B)}{Re(B)}\right)$. At antinodes the number of incident electrons available for tunneling decreases and maximum tunneling current density thus correspond to the spacer layer width which forms a node at the end of the spacer layer. All these observations point at maximum PVCR for a spacer layer width of about half the De Broglie wave length corresponding to the resonant energy. To sum up, the optimum well barrier and spacer layer width for the the GaAs –Ga$_{0.7}$Al$_{0.3}$As RTD amounts to 38 Å, 47 Å and 62 Å respectively and the corresponding highest PVCR and the peak current amounts to 229 and 16.20kA/cm$^2$.

As regards experimental results reported, a peak current of 270 kA/cm$^2$ was measured in a 51Å Ga$_{0.7}$Al$_{0.3}$As- 31Å GaAs –51Å Ga$_{0.7}$Al$_{0.3}$AsRTD [31] . Soderstrom et al. [32] reported the current voltage relation for a 25ÅGa$_{0.7}$Al$_{0.3}$As- 45ÅGaAs –25ÅGa$_{0.7}$Al$_{0.3}$As TD with a spacer layer width of 25Å at room temperature and did not report any peak in the relation. For these parameters we also did not find any peak in the J~V relation. The PVCR of value 1.85 obtained by Jogai et al. [33], we obtained a PVCR of 1.47 for the same structure at 300K which validates our computational method and the obtained results.

The graph in Fig. 5 points to an increase in PVCR with increase in spacer width decreasing after reaching a certain PVCR.

The graphs in Fig. 3, 4 and ?? point that the RTD made of different semiconductor materials characterized by the barrier height and effective mass parameters favor an optimum device dimension specially the well, barrier and spacer layer width wherein the PVCR is found to be highest. As such, we make an effort to find the optimum parameters for highest PVCR in the GaN –Ga$_{0.7}$Al$_{0.3}$N RTD.

**Optimum device parameters in GaN –Ga$_{0.7}$Al$_{0.3}$N RTD**

Here we have made an effort to find the optimum device parameters corresponding to highest PVCR in a system akin to GaN –Ga$_{0.7}$Al$_{0.3}$N. The barrier height and the effective mass of the well and barrier materials are taken as 0.5111eV, 0.2000m$_0$ and 0.2343m$_0$ respectively. The J-V characteristics are calculated at the temperature of 300K with the system parameters $\varepsilon_F$ = 0 i.e. the Fermi level coinciding with the conduction band edge of GaN corresponding to a doping of 0 .57%. In Fig. 6(a) we present the variation of PVCR with changes in well width. It is worth pointing here that the barrier width corresponding to each well width is chosen such that the PVCR gets maximized. The well and barrier width for achieving optimum PVCR in the system amounts to 21Å and 32Å respectively. The optimum width of the spacer layer for the system amounts to 35 Å. It is worth pointing here that the bound energy level of the corresponding quantum well amounts to 0 .1556eV and the de Broglie wavelength for the electron stands at 69.5 Å. The optimum width of the spacer layer equals half of the de Broglie wavelength.

Under the specified parameter conditions the PVCR amounts to 2128.39 with a peak current density of 7.584 kA/cm$^2$. The value of peak current density is comparable to the reported value of 14kA/cm$^2$ in a 24 Å GaN –24Å Ga$_{0.82}$Al$_{0.18}$N-24 Å GaN RTD with a 20Å GaN spacer layer [34] .The width of the NDC region is found to be 0.0421V. The J-V graph under the optimum conditions is presented in Fig. 6(b). As expected with increase in barrier height in GaN/ Ga$_{0.7}$Al$_{0.3}$N system than that of the GaAs –Ga$_{0.7}$Al$_{0.3}$As system, though there is a decrease in the peak and valley current, the PVCR is quite higher.

**Relation of optimum device parameters on barrier height and effective mass**

As discussed in the earlier sub section, there exists a unique set of device parameters for each of the GaAs –Ga$_{0.7}$Al$_{0.3}$As and GaN –Ga$_{0.7}$Al$_{0.3}$N RTD system for which the PVCR is the highest. The material properties of different RTD system appear in the current density calculation through the effective mass, the dielectric constant and the conduction band mismatch i.e., the barrier height, of the host materials. We thus felt the need to find a relation of the optimum device parameters (well and barrier width) in different systems on the barrier height and the effective masses of the well and barrier materials

In Fig.7, we have presented the variation of optimum well and barrier width required to achieve highest PVCR vs. barrier potential for two sets of different effective masses. From the figure, it can be ascertained that with the increase in barrier potential the optimum well width decreases and the corresponding barrier width increases almost linearly. Further, for the barrier height of 0.23eV, optimum PVCR is obtained under equal width of the well and barrier [$a = b = 41.5$ Å for $m_w^* = 0.067m_0$ and $m_b^* = 0.0904m_0$; $a = b = 25$ Å for $m_w^* = 0.20m_0$ and $m_b^* = 0.23m_0$]. The figure also points at an increase in the optimum well and barrier width with a decrease in effective mass of the materials.

The variation of well width with the inverse square root of effective mass of well material $(m_W^*)^{-1/2}$ and barrier width with the inverse square root of effective mass of barrier material $(m_B^*)^{-1/2}$ are found to exhibit almost linear relationship as shown in Fig 8(a) and (b). The linear graphs indicate that the well width and the barrier width are inversely proportional to the square root of the effective mass associated to their material. It is worth pointing here that in the quantum confinement along the growth direction the well width should be in the range of the De Broglie wavelength and as such is inversely proportional to the square root of the effective mass of the well material.

**Table. IV.** Data corresponding to the Fig 8.

| $V_0$ (eV) | $m_w^*$ ($m_0$) | $m_B^*$ ($m_0$) | $a_o$ (Å) from J~V relation | $b_o$ (Å) from J~V relation | $a_o$ (Å) from relation (7) | $b_o$ (Å) from relation (8) |
|---|---|---|---|---|---|---|
| 0.23 | 0.067 | 0.0904 | 41.5 | 41.5 | 41.3 | 41.3 |
| 0.3728 | 0.067 | 0.0904 | 38 | 47 | 38 | 47.1 |
| 0.5111 | 0.067 | 0.0904 | 35 | 53 | 34.8 | 52.7 |
| 0.23 | 0.2 | 0.2343 | 25 | 25 | 24.8 | 25.4 |
| 0.3728 | 0.2 | 0.2343 | 23 | 29 | 22.6 | 29.1 |
| 0.5111 | 0.2 | 0.2343 | 21 | 32 | 20.4 | 32.6 |
| 0.23 | 0.32 | 0.36 | 20 | 20 | 20 | 20.4 |
| 0.3728 | 0.32 | 0.36 | 18 | 23 | 18.1 | 23.4 |
| 0.5111 | 0.32 | 0.36 | 16 | 26 | 16.3 | 26.3 |

Further the optimum spacer layer is found to be nearly half of the de Broglie wave length associated with the bound state of the corresponding finite quantum well of optimum width. The data corresponding to the Fig 8 is tabulated in Table IV for the purpose of finding the relations for the optimum well width and the barrier width in terms of $V_0$ and $m_w^*$ or $m_B^*$. The well width and barrier width were found to have relations:

$$a_o = (11.2 - 4.6V_0)\left(m_w^*/m_0\right)^{-\frac{1}{2}} + (3.3 - 5.2V_0) \quad (7)$$

$$b_o = (9.8 + 12V_0)\left(m_B^*/m_0\right)^{-\frac{1}{2}} - (0.7 - 0.8V_0) \quad (8)$$

The values of the optimum well width and the barrier width computed on the basis of the (7) and (8) as provided in the last two columns of the table tally well with those obtained from the computation of current density versus applied potential as provided in the 4$^{th}$ and 5$^{th}$ column of the Table. V.

**Effect of barrier height and effective mass on highest PVCR:**

The variation of highest PVCR corresponding to the optimum parameters with respect to change in the effective mass of well, for two different values of $V_0$ is presented in Fig. 9. The figure shows that for a constant barrier height, the PVCR decreases slightly with the increase in effective mass. However, with the increase in barrier height the PVCR increases appreciably.

**Effect of doping concentration on PVCR and peak current density for optimum device parameters**

The effect of doping concentration on the PVCR and the peak current density is being presented in Fig. 10(a) and (b) and its insets respectively, for the 38Å GaAs –47Å$Ga_{0.7}Al_{0.3}$As with a spacer layer of 62Å and the 21Å GaN/32Å $Ga_{0.7}Al_{0.3}$N with 35Å spacer layer RTDs. As pointed in the numerical analysis section, the doping concentration in the contact layer appears in the current density through the Fermi energy. To comprehend clearly the effect of doping concentration, for that matter the Fermi energy, on the device parameters, we have listed in Table IV, the values of peak current and the PVCR corresponding to different doping concentrations in the GaAs –$Ga_{0.7}Al_{0.3}$As RTD under optimum device parameter condition.

**Table. V.** Peak current density, peak voltage and other associated parameters for the 38Å GaAs – 47Å$Ga_{0.7}Al_{0.3}$As RTD with different doping concentration in the contact layer.

| Doping concentration (atoms/cc) | Fermi energy (eV) | Peak current density (kA/cm$^2$) | PVCR |
|---|---|---|---|
| $5 \times 10^{17}$ | -0.0149 | 10.40687 | 250.1324 |
| $10^{18}$ | -0.0027 | 15.02139 | 233.6845 |
| $1.17 \times 10^{18}$ | 0 | 16.21922 | 229.5356 |
| $5 \times 10^{18}$ | 0.0246 | 30.05938 | 185.1186 |
| $10^{19}$ | 0.0364 | 38.35747 | 161.5177 |

**Resonant Tunneling Lifetime and Operating Frequency**

Resonant tunneling lifetime $\tau$ determined from the data of Fig. 2(b) for the 38Å GaAs–47Å $Ga_{0.7}Al_{0.3}$As system amounts to $1.56 \times 10^{-12}$s and a corresponding operating frequency of around $0.64 \times 10^{12}$ Hz. The corresponding lifetime and operating frequency in the 21Å GaN–32 Å $Ga_{0.7}Al_{0.3}$N system were found to be $1.265 \times 10^{-11}$s and $0.79 \times 10^{11}$ Hz respectively. The operating frequency in the GHz range makes the RTD most desirable for applications in high frequency oscillators. The GaAs/ $Ga_{0.7}Al_{0.3}$As with higher peak current and high power seems to be more suitable for analog circuits while GaN/ $Ga_{0.7}Al_{0.3}$N RTD with improved PVCR is a better candidate for digital circuits.

**Conclusion**

For a particular RTD based on two host materials characterized by a constant barrier height and constant effective masses, corresponding to a well width there exist a characteristic barrier width where the PVCR become maximum. For example in GaAs/Ga$_{0.7}$Al$_{0.3}$As RTD for the well width of 20Å, 25 Å and 50Å, the maximum PVCR is obtained for a barrier width of 46 Å, 53 Å and 41 Å respectively. The barrier width corresponding to maximum PVCR occurs where the resonant tunneling energy of the DBS becomes equal to the bound energy of the quantum well.

With an increase in well width, the maximum PVCR first increases attains a maximum and then decreases indicating the existence of a unique well and barrier width for which the PVCR is the highest. For example the optimum well and barrier width for maximum PVCR in GaAs/Ga$_{0.7}$Al$_{0.3}$As and GaN/Ga$_{0.7}$Al$_{0.3}$N RTDs correspond to 38Å, 47Å and 21Å, 32Å, respectively.

The optimum spacer layer corresponds to nearly half of the de Broglie wavelength associated with the bound state of the corresponding finite quantum well. For the two systems studied the width of spacer layer is found to be 62Å and 35 Å, respectively.

The PVCR is found to increase appreciably with an increase in barrier height. At room temperature with the Fermi energy coinciding the conduction band edge of well material, the highest PVCR for the GaAs/Ga$_{0.7}$Al$_{0.3}$As and GaN/Ga$_{0.7}$Al$_{0.3}$N based RTD under optimum device parameter condition is found to be 229.5 and 2128.4, respectively.

The well width and barrier width corresponding to highest PVCR is found to depend linearly on the inverse square root of the effective mass of the well and barrier material respectively. The optimum well width and barrier width for any RTD based on the effective mass of the well and barrier materials and the barrier height at the hetero junction can be calculated on the basis of the relations (7) and (8). The optimum width of spacer layer being equal to half the De-Broglie wavelength associated with the bound state energy of the finite well can be calculated as detailed in Appendix-1. The relations developed will provide the experimentalists a tool to calculate the optimum device parameters for RTDs based on any two materials to achieve highest PVCR and our calculations for both the systems under consideration showed that they oscillate in THz frequency range.

## Appendix-A

### Energy and wave function in a Quantum Well

In Fig. A1, we have presented the potential profile in black solid line for the finite quantum well. The associated Schrodinger equation can be solved and the energy level can be obtained by employing the continuity of probability density obtained through $\psi_W(x) = \psi_B(x)$ and the continuity of current density expressed through $\frac{1}{m_W^*}\frac{\partial \psi_W(x)}{\partial x} = \frac{1}{m_B^*}\frac{\partial \psi_B(x)}{\partial x}$ at the well barrier junctions [at $x = -a/2$ and $a/2$]. The notations, $\psi_W(x)$ and $\psi_B(x)$ represent the envelope function for the well and barrier regions respectively. The lowest bound state energy of the well is computed from the relation

$$\frac{ka}{2} \tan \frac{ka}{2} = \gamma \frac{\beta a}{2} \tag{A1}$$

where, $k = \sqrt{\frac{2m_W^* \varepsilon}{\hbar^2}}, \beta = \sqrt{\frac{2m_B^*(V_0 - \varepsilon)}{\hbar^2}}$ and $\gamma = \frac{m_W^*}{m_B^*}$

The envelope function corresponding to the energy level is computed using the relations:

$$\psi_W(x) = A \cos kx \quad \text{and} \quad \psi_B(x) = A \cos \frac{ka}{2} e^{-\beta\left|x - \frac{a}{2}\right|} \tag{A2}$$

The bound state energy for the Ga$_{0.7}$Al$_{0.3}$As 38ÅGaAs/Ga$_{0.7}$Al$_{0.3}$As finite quantum well computed on the basis of (A1) amounts to 0.12349 eV. The probability density of the electron in the bound state computed on the basis of (A2) along with the energy level is shown in the Fig. (A1).

The bound state energy for the Ga$_{0.7}$Al$_{0.3}$N- 21ÅGaN/Ga$_{0.7}$Al$_{0.3}$N finite quantum well computed on the basis of (A1) amounts to 0.1557 eV.

## Appendix –B

The energy band gaps of the mixed compounds are obtained using the relations:

$\varepsilon_g$(Ga$_{1-x}$Al$_x$ As) = (1-x) × $\varepsilon_g$ (GaAs) + x× $\varepsilon_g$ (AlAs)-x × (1-x) × (1.31× x-0.127)

And $\varepsilon_g$(Ga$_{1-x}$Al$_x$ N) = (1-x) *$\varepsilon_g$ (GaN) + x*$\varepsilon_g$ (AlN) - x*(1-x)*1.00(17)

The barrier height $V_0$, which is equal to the hetero junction conduction band offset, is taken as 88% of the difference between the energy gaps of the two host materials. In both Ga$_{0.7}$Al$_{0.3}$As and Ga$_{0.7}$Al$_{0.3}$N compounds, the band structure is considered to be GaAs and GaN like respectively, and their effective masses are obtained using the relation

$$\frac{m^*_{GaAs}}{m^*_{GaAl\,As}} = \frac{\varepsilon_{g_{GaAs}}}{\varepsilon_{g_{GaAlAs}}}$$

The Fermi energy of the two systems is computed for the doping concentration and the donor ionization energy using the relations

$$n = N_c F_{\frac{1}{2}}\left(\frac{\varepsilon_F}{k_B T}\right) = \frac{N_D^+}{2} + \sqrt{\left(\frac{N_D^+}{2}\right)^2 + n_i^2} \text{ and } N_D^+ = N_D \left(1 + 2e^{\frac{(\varepsilon_F - \varepsilon_D)}{k_B T}}\right)^{-1}$$

where, $n$, $n_i$, $N_c$, $F_{\frac{1}{2}}$, $N_D$ and $N_D^+$ represent the electron concentration, intrinsic carrier concentration, effective conduction band density of states, and Fermi Dirac integral of order ½, donor concentration and ionized donor concentration respectively.

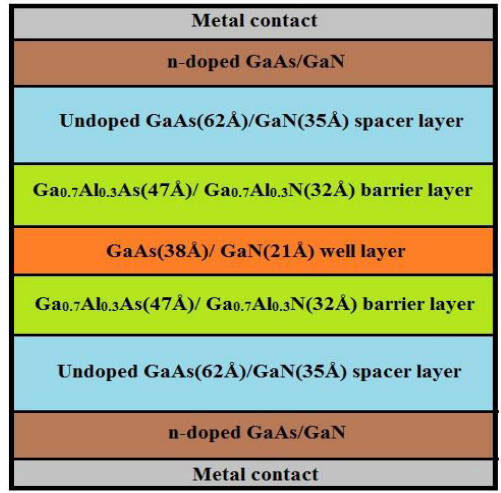

**Fig. 1:** A schematic layer structure of the RTD

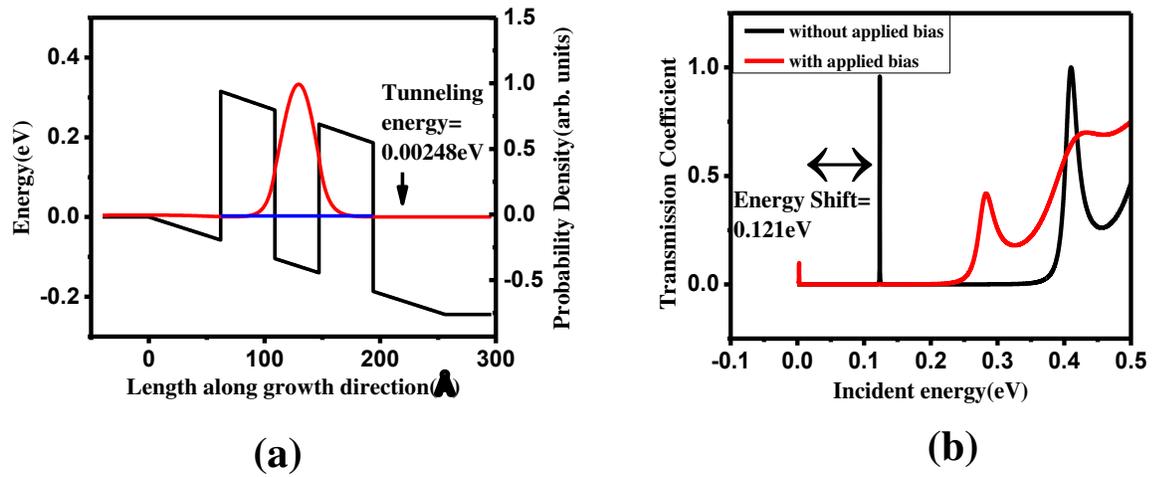

**Fig. 2: (a)** The potential profile (black) along with probability density (red) and resonant tunneling energy(blue) of a double barrier 47Å$Ga_{0.7}Al_{0.3}As$/ 38ÅGaAs /47Å$Ga_{0.7}Al_{0.3}As$ system in the presence of a bias potential of 0.2444V. **(b)** Transmission coefficient ~ Incident energy plot in both the absence (black) and the presence (red) of applied bias (0.2444V). The first resonant tunneling peak in these cases correspond to 0.12348eV and = 0.00248eV respectively. The shift in the RTE due to applied bias = 0.121eV.

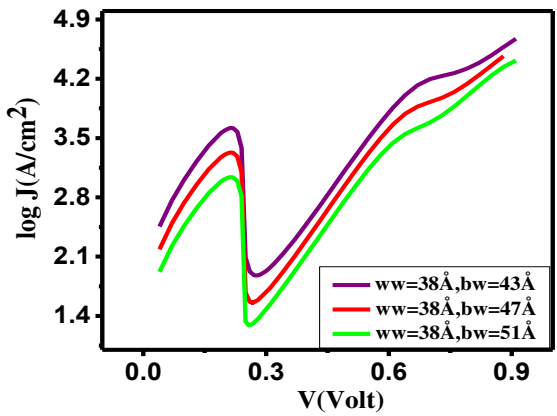
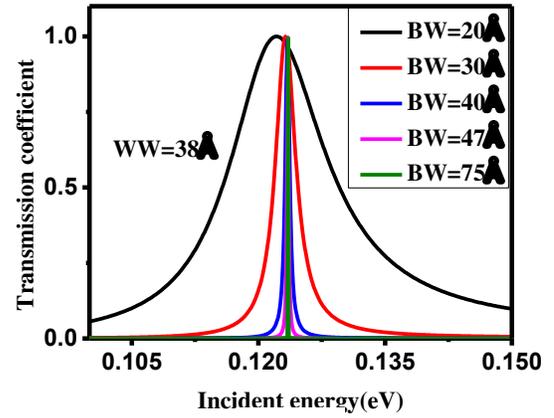

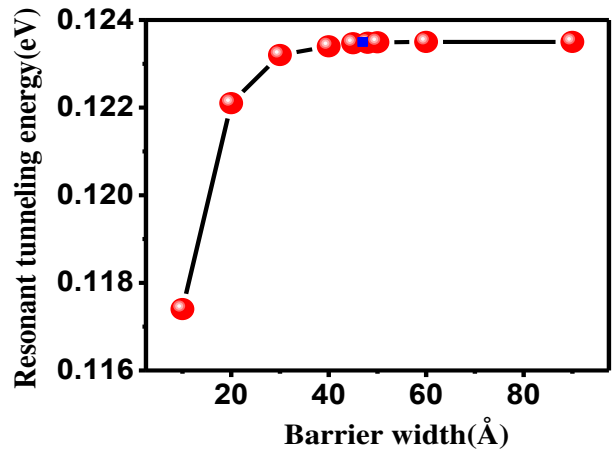

**Fig. 3:** **(a)** J~V Characteristic of GaAs/Ga$_{0.7}$Al$_{0.3}$As RTD for different barrier widths from 43Å to 51Å keeping well width constant at 38 Å at 300K and $\varepsilon_F = 0$. **(b)** The transmission coefficient and the resonant tunneling energy for various values of the barrier width in the GaAs – Ga$_{0.7}$Al$_{0.3}$As RTD structure keeping the well width at 38Å. **(c)** Variation of resonant tunneling energy with barrier widths in GaAs/Ga$_{0.7}$Al$_{0.3}$As DBS with well width of 38 Å. For barrier width ≥ 47Å the resonant tunneling energy of DBS become equal to the bound state energy of the corresponding quantum well.

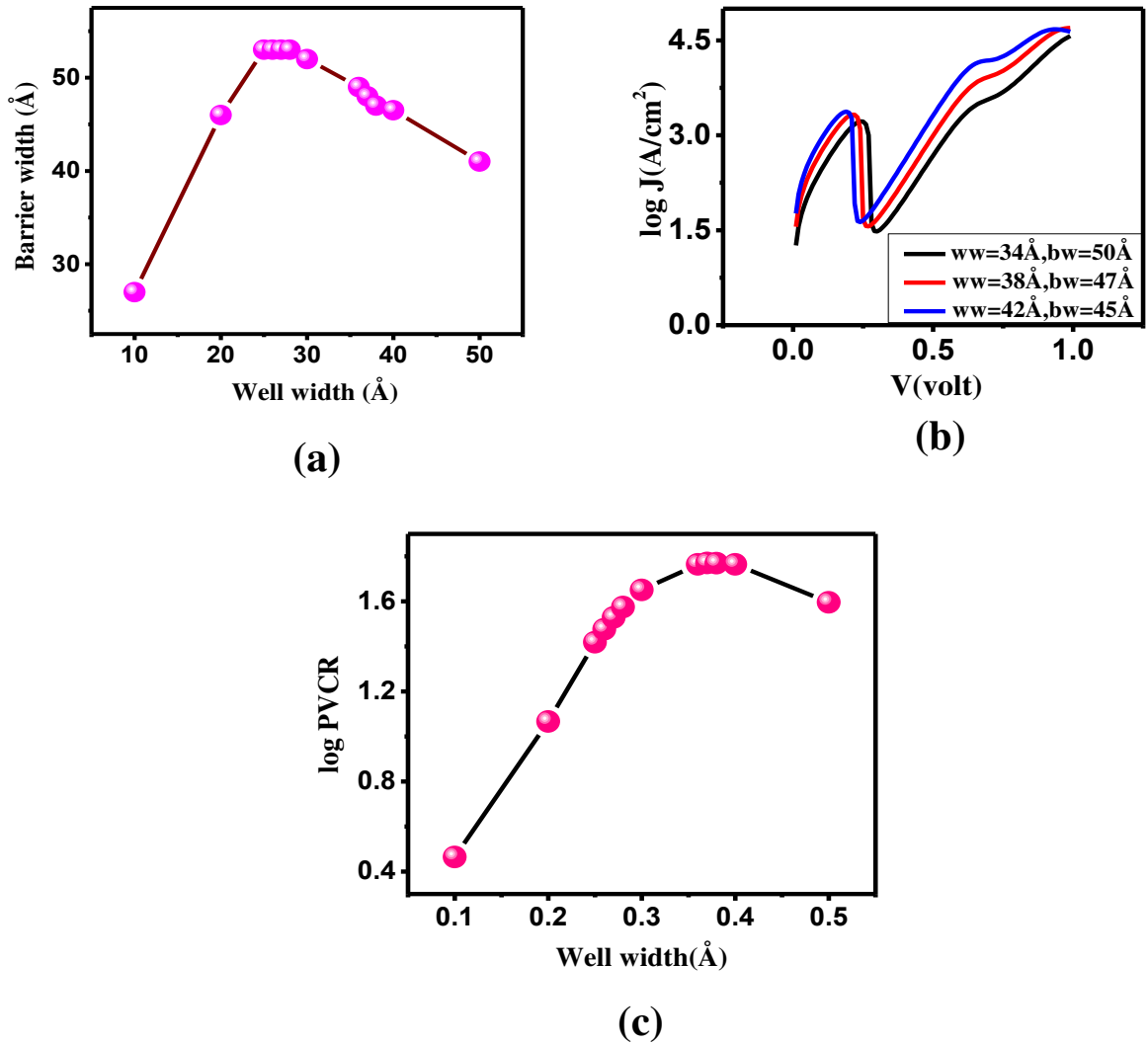

**FIG. 4: (a)** Variation of barrier width with well width to achieve maximum PVCR. **(b)** J~V Characteristic of GaAs/Ga$_{0.7}$Al$_{0.3}$As RTD for various well width with the barrier width wherein the PVCR is maximum. **(c)** PVCR~ Well width for GaAs/Ga$_{0.7}$Al$_{0.3}$As RTD showing the optimum PVCR at 38Å well width and 47Å barrier width.

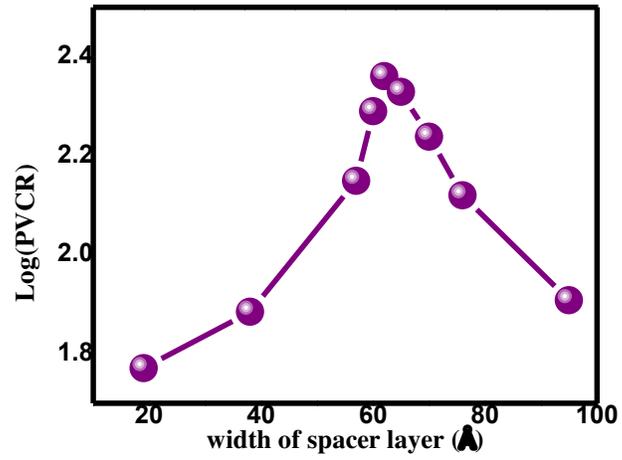

**Fig. 5:** Variation of PVCR with changing widths of spacer layers indicating a maximum PVCR for a certain spacer layer width of 62Å for our proposed model of GaAs/ Ga0.7Al0.3As.

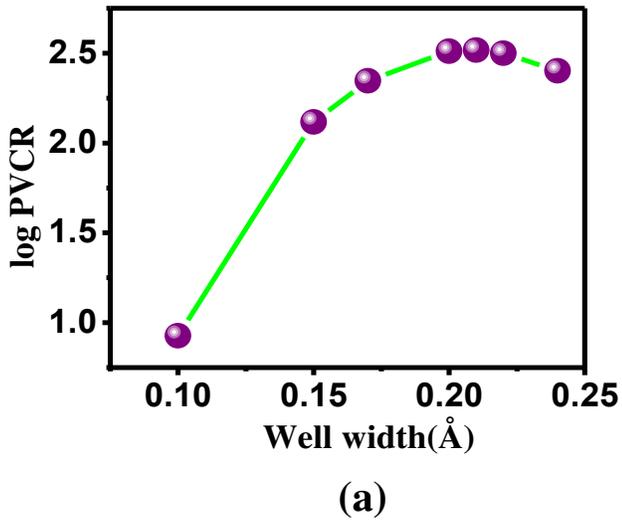 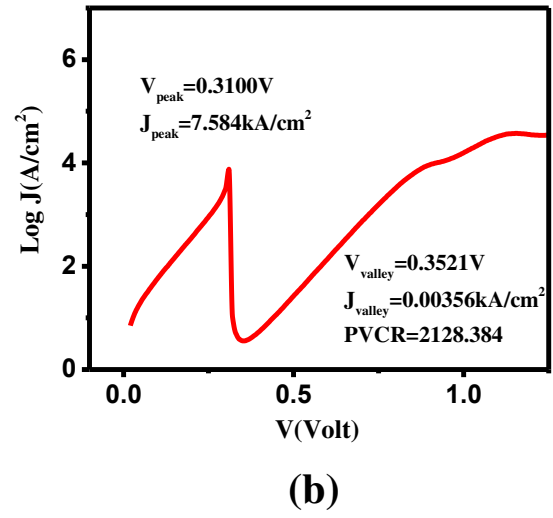

**Fig. 6:** **(a)** PVCR~ well width of GaN/Ga$_{0.7}$Al$_{0.3}$N system showing optimum values for wellwidth of 21Å and barrier width of 32Å. **(b)** J~V characteristic of 21Å GaN/32Å Ga$_{0.7}$Al$_{0.3}$N system at room temperature and $\varepsilon_F = 0$ with a spacer layer of 35 Å.

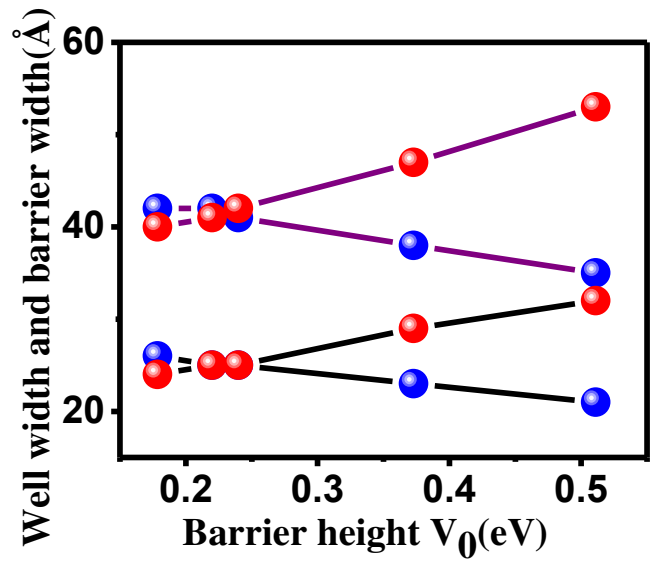

**Fig. 7:** Well width and barrier width~ barrier height corresponding to optimum PVCR for two different effective masses of well and barrier materials. The black solid/dash lines show variation of well width and barrier width with barrier height for a set of effective masses $m_w^* = 0.2000m_0$, $m_b^* = 0.2343 m_0$ and the purple coloured solid /dash lines show the same for a different set of effective masses i.e $m_w^* = 0.0670m_0$, $m_b^* = 0.0904m_0$.

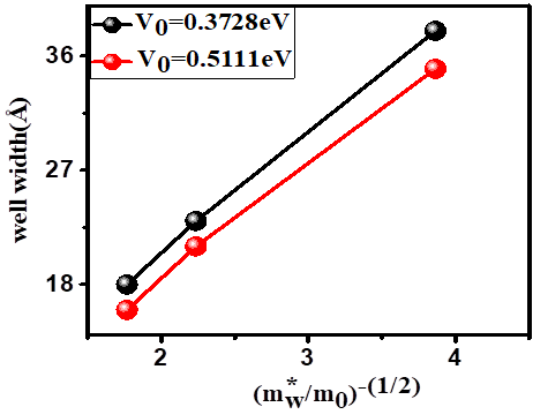 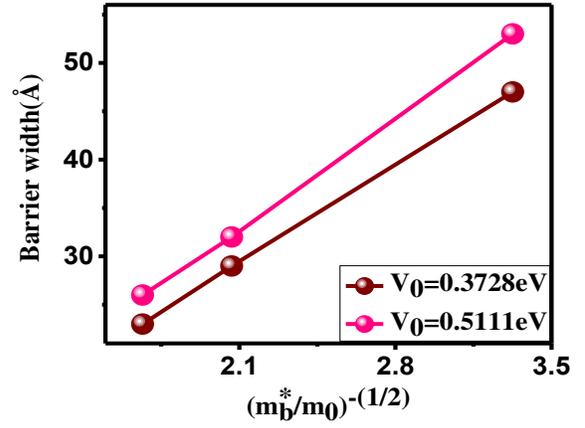

**Fig. 8:** Variation of **(a)** $(m_w^*)^{-1/2}$ with well width and **(b)** $(m_b^*)^{-1/2}$ with barrier width for two different constant barrier heights.

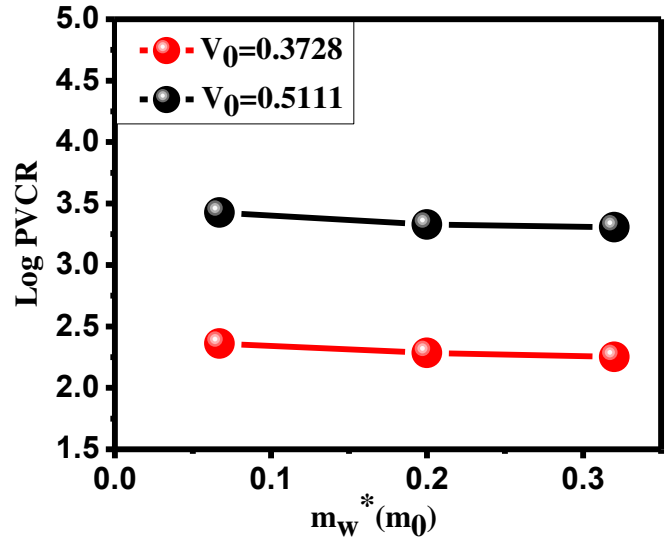

**Fig. 9:** PVCR~ effective mass of the well ($m_w^*$) for two different barrier heights

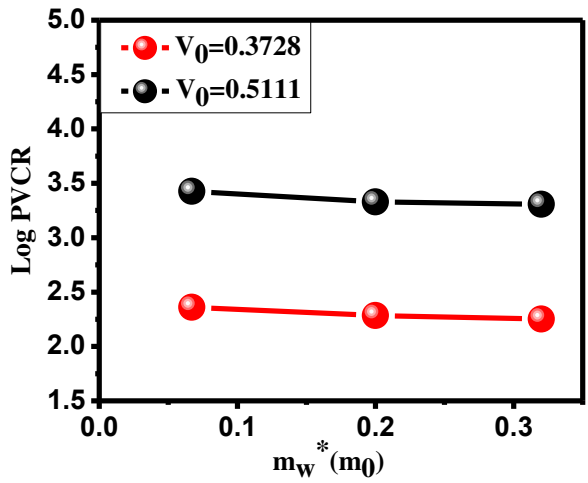 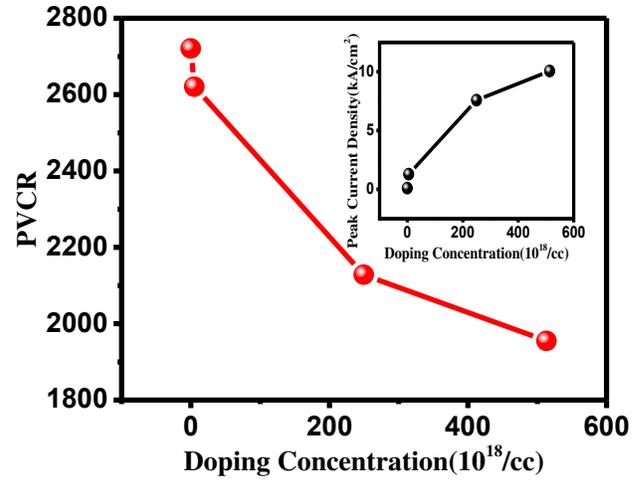

**Fig. 10:** **(a)** Variation of PVCR with various doping concentrations and the inset Peak current ~ doping concentrations for GaAs/Ga$_{0.7}$Al$_{0.3}$As RTD. **(b)** Similar plot for GaN/Ga$_{0.7}$Al$_{0.3}$RTD.

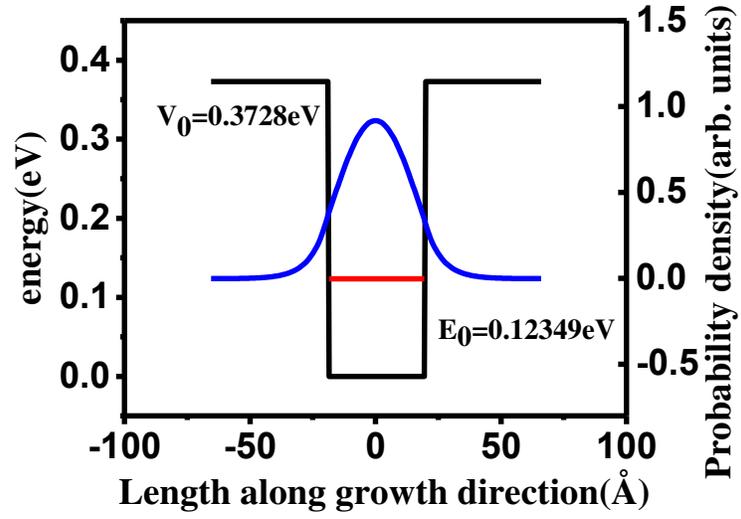

**Fig. A1:** Potential profile of the finite quantum well formed from $Ga_{0.7}Al_{0.3}As$ 38Å$GaAs$/$Ga_{0.7}Al_{0.3}As$ is presented in Black solid line. Blue line represents the probability density of the electron in this state. Red line indicates the bound state energy of the well ($E_0$).